# Space Weather Prediction with Exascale Computing

Giovanni Lapenta (giovanni.lapenta@wis.kuleuven.be)

Centrum voor Plasma-Astrofysica, Departement Wiskunde,

Katholieke Universiteit Leuven, Belgium (EU).

Space weather refers to conditions on the Sun, in the interplanetary space and in the Earth space environment that can influence the performance and reliability of space-borne and ground-based technological systems and can endanger human life or health. Adverse conditions in the space environment can cause disruption of satellite operations, communications, navigation, and electric power distribution grids, leading to a variety of socioeconomic losses. The conditions in space are also linked to the Earth climate. The activity of the Sun affects the total amount of heat and light reaching the Earth and the amount of cosmic rays arriving in the atmosphere, a phenomenon linked with the amount of cloud cover and precipitation. Given these great impacts on society, space weather is attracting a growing attention and is the subject of international efforts worldwide.

We focus here on the steps necessary for achieving a true physics-based ability to predict the arrival and consequences of major space weather storms. Great disturbances in the space environment are common but their precise arrival and impact on human activities varies greatly. Simulating such a system is a grand- challenge, requiring computing resources at the limit of what is possible not only with current technology but also with the foreseeable future generations of super computers

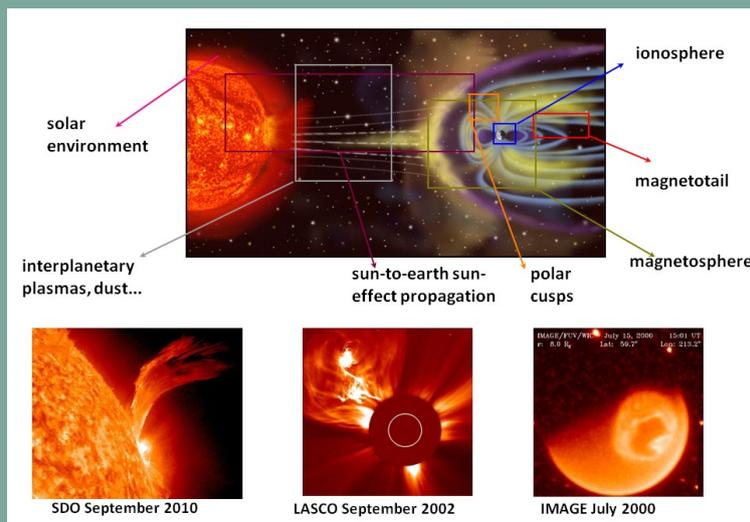

Figure 1: Different phases of a space weather event.
Top: artist view from NASA
Bottom: images from different past space missions.

The Sun emits a constant solar wind in the form of a hot ionized gas and a magnetic field. The Solar wind interacts with the Earth magnetic field forming an elongated magnetosphere that protects the Earth environment from many of the most devastating consequences. Strong perturbations and high energy particles still penetrate creating serious trouble in space and on Earth

## Space Weather and the Exascience Lab

The actors of space weather are the Sun, the Earth and the vast space in between. Like any star, the sun is made of a highly energetic and highly conductive gas, called plasma. In plasmas, the atoms have been broken into their nuclei and their electrons that become free to move. The hot plasma of the Sun is confined by gravity and moves in complex patterns that produce large currents and large magnetic fields. The gravitational confinement is not perfect and a highly varying outflow of plasma, called solar wind, is emitted from the Sun to permeate the whole solar system, reaching the Earth. The Earth has itself a magnetic field. The Earth magnetic field makes compasses point towards the north pole and allows many species of animals to find their way during migrations. This very same field protects the Earth from the incoming solar wind and its disturbances. Only a small part of the particles can reach the Earth surface, the so-called cosmic rays. Most of the incoming plasma is stopped and deflected, reaching only high strata of the atmosphere at the polar regions and causing the aurora seen by the peoples of the northern latitudes as the northern lights.

Space weather simulations must follow the plasma emitted from the Sun and track its evolution in the interplanetary space, focusing on its impact on the Earth. The different phases of the processes are described in Fig.1

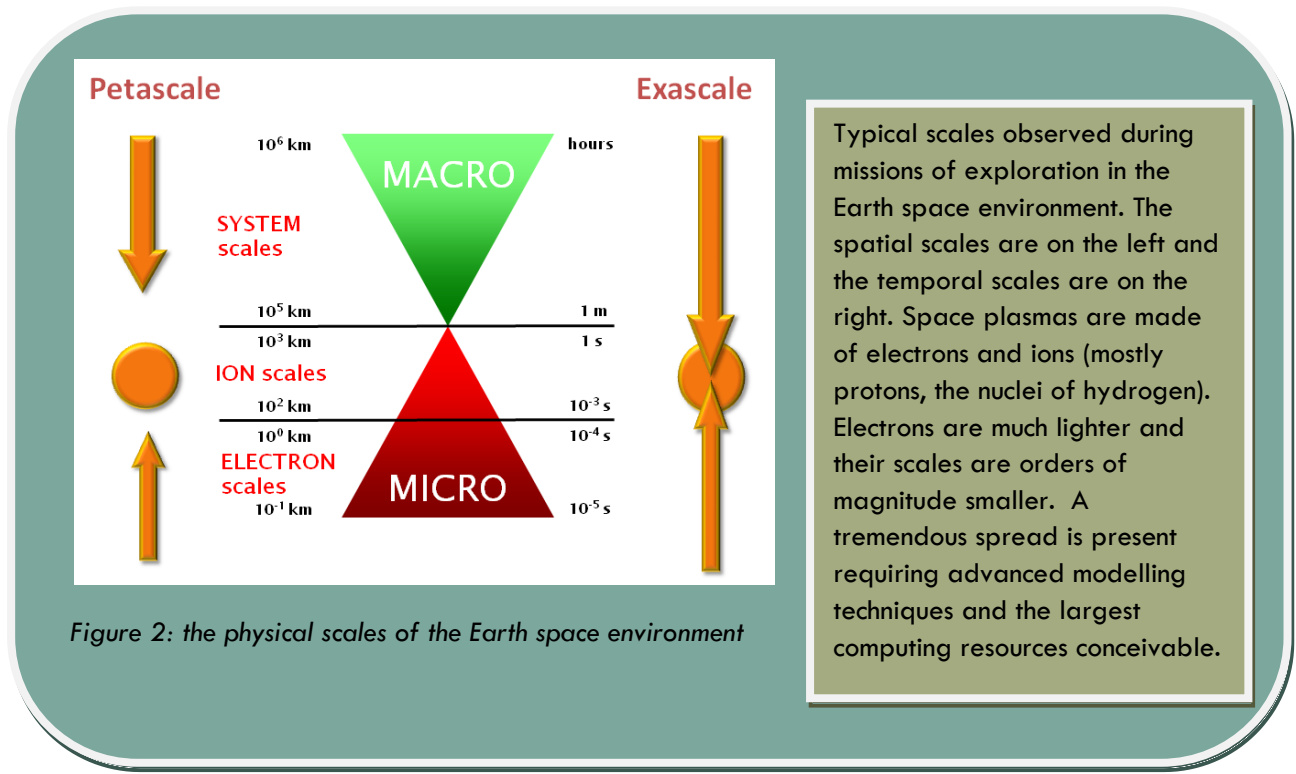

*Figure 2: the physical scales of the Earth space environment*

Typical scales observed during missions of exploration in the Earth space environment. The spatial scales are on the left and the temporal scales are on the right. Space plasmas are made of electrons and ions (mostly protons, the nuclei of hydrogen). Electrons are much lighter and their scales are orders of magnitude smaller. A tremendous spread is present requiring advanced modelling techniques and the largest computing resources conceivable.

To model such processes, both electromagnetic fields and plasma particles need to be described. The nuclei (mostly protons, the nuclei of hydrogen) and the electrons of the plasma are coupled together loosely and each species has its own typical scales. The electromagnetic fields keep the species coupled forming a very non-linear and multi-scale system.

Modeling space weather is a daunting task: daunting because the system is enormous and because it includes a wide variety of physical processes and of time and space scales. Fig. 2 shows the typical scales observed from space exploration missions in the Earth environment. The scales are presented in the form of a hourglass with the top part presenting the macroscopic scales of evolution and at the bottom the smallest scales of importance.

A computer model of space weather must face this challenge by using state of the art mathematical techniques to deal with non-linear multiscale systems. The Exascience Lab relies on the implicit moment method, intuitively described in Fig. 3. The fields and the particles are studied together in a coupled manner. The word implicit refers to the ability of the method to advance both fields and particles together without any lag between the two (the time lag is a typical aspect of the explicit methods, instead). The word moment refers to the use of moments of the particle statistical distribution. The moments are local statistical averages that characterize the particle properties.

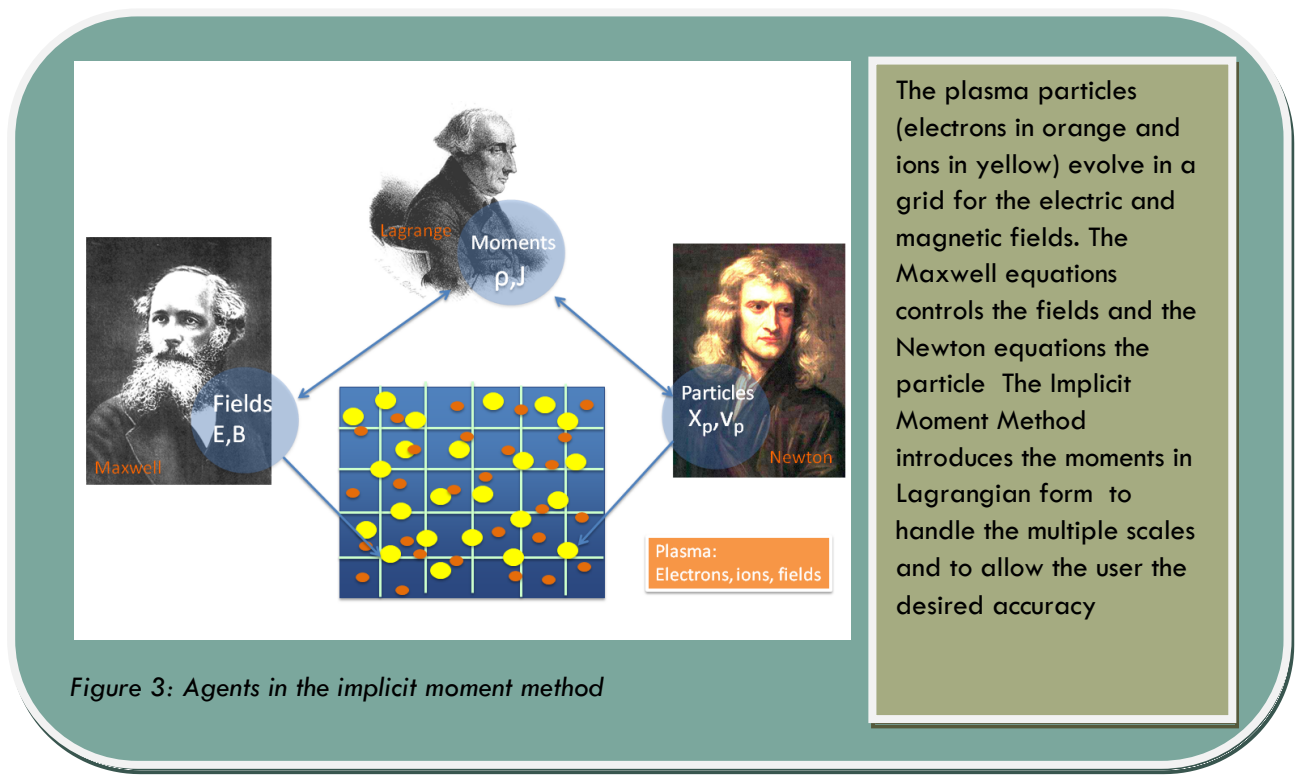

*Figure 3: Agents in the implicit moment method*

The implicit moment method has been developed at Los Alamos by Jerry Brackbill in the 80's and the author has been the leader of the development effort at Los Alamos during the last decade, before moving to the present effort. At Leuven, the new code iPic3D has been developed by Stefano Markidis and the author and is in use and under further development by the Exascience Lab.

The implicit moment method allows us to select the local level of resolution according to the scales of the local processes. This feature allows to model space weather events with the minimum effort, increasing the resolution only where absolutely needed. Even with this method, however, the current petascale computers can only model subsections of a typical space weather event. With existing resources, reduced models need to be used. Some of the processes are neglected and approximated with adjustable parameters and heuristic assumptions, reducing the predictive value of the approach. To achieve a truly predictive model of space weather events exascale computing will be needed.

## Space Weather Modelling on Exascale

Figure 4 illustrates the current situation of the state of the art in space weather modeling. The full description based on physically sound first principles requires one to resolve the electron scales in very localized regions of the system where energy conversion processes develop. In the figure, we report a state of the art simulation based on existing limited models that do not capture the full physics. To reach truly predictive capability exascale resources will be needed.

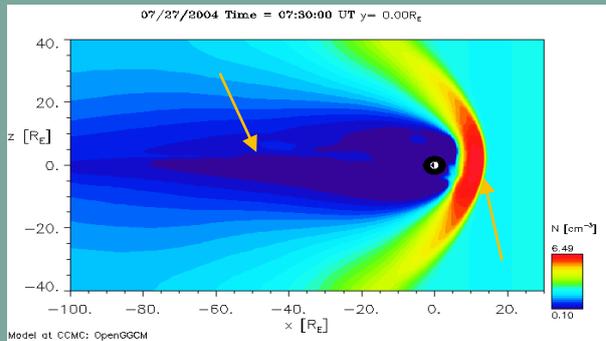

| METHOD | CELLS PER DIMENSION | CORES |
|---|---|---|
| Explicit Approach | 6,353,000 | 6.2600e+16 |
| Implicit Moment Method | 63,530 | 6.2600e+10 |
| Implicit Moment Method with AMR | Coarse: 635 Adaptivity on localized regions | Coarse level: 6.2600e+04 Refinement: 1.9812e+06 |
| On petascale (order 100,000 cores) supercomputers, it is not possible to simulate space weather events with full resolution even with the most advanced modelling method. Exascale will instead allow a fully resolved model. | | |

*Figure 4: Physics-based model of space weather.*

*Cut of a 3D simulation with OpenGGCM done by the Author at the facilities of the CCMC at NASA Goddard. The regions requiring additional resolution are identified with arrows. Only a very thin layer in ach place needs to be resolved.*

A simple calculation provides the reason. Missions of exploration and models provide a clear indication of the scales involved (see Figs. 1-2). The needed simulation box for the modeling the Earth environment must be of the order of 100 Earth radii in each of the 3 spatial direction ($100R_E$ x $100R_E$ x $100R_E$) where the Earth radius is 6353Km. As observed in Fig.2, the smallest spatial scales are the electrostatic responses (called Debye length) at about 100m. Clearly simulating the whole box of size of more than 600 thousand kilometers with a resolution of 100 meters will require impossible resources (for the foreseeable future). Yet that is exactly what the most common methods in use would need to do. The so-called explicit methods need to resolve the smallest electrostatic responses and the table in Fig. 4 reports the impractical requirements of such a calculation. Our experience leads us to assume that a single core of current design can fit at most 16x16x16 cells each with a hundreds of particles for each species (electrons and ions). With that assumption a current state of the art explicit code would require 63 millions of billions of cores. Not even exascale can provide that.

At the exascience lab, however, the focus is on implicit moment methods that allow to select the local accuracy. Two steps can then be followed. First, just the use of the implicit method allows the user to avoid the need to resolve the electrostatic scales. Instead, the method averages over them and can allow the resolution to be increased to the electron electromagnetic response (electron inertial range), at about 10Km. This mathematical modeling approach saves then 2 orders of magnitude in each direction and similarly allows a corresponding reduction of time steps needed. The user can rise the bar on the hourglass of Fig. 2 by two notches in space and in time. Keeping the bar horizontal, the implicit moment method allows to rise the resolution in space and in time concurrently. A uniform grid with the implicit moment method still requires 63 billions of cores. This is closer to being within reach but it is still too large a number. This resolution remains still outside the reach of even future exascale computers.

A second and final step is taken in our approach: adaptive grids. The regions of dissipation where the simulations truly need to resolve the electron dynamics happen on very limited regions (see Fig. 4) of reduced dimensionality: thin crusts over surfaces dividing different types of plasma (for example the solar wind from the Earth magnetosphere). A reasonable assumption guided by practical experience is to use a layer of 10 cores away from the surfaces of interest. Additionally, we assume the impacted surfaces to be about 10% of a sphere encircling the Earth at the distances of interaction with the solar wind, an estimate can be computed of the total number of cores needed. For the large scale modeling of the wholes system, resolving just the ion scales, a very manageable number of cores is needed: 62 thousand. This is well within reach of petascale computing and indeed currently this is the state of the art. But to resolve the electron physics and avoid ad hoc assumptions that limit the predictive value of the simulations, the AMR approach will need an additional 1.98 million cores. All together the first ever predictive simulation based on physics principles will require about 2 million cores and will be feasible on the first exascale computer.

---